\documentclass{article}

\pdfoutput=1
\usepackage{fullpage}
\usepackage[utf8]{inputenc}
\usepackage{graphicx}
\usepackage{amssymb}
\usepackage{amsmath}
\usepackage{amsfonts}
\usepackage{color}
\usepackage[pdftex]{hyperref}
\usepackage{hyperref}
\usepackage[english]{babel}
\usepackage{subfigure}
\usepackage{algorithm}
\usepackage{algorithmic}
\usepackage{multicol}
\usepackage{multirow}
\usepackage{tabularx,colortbl}
\usepackage{lineno}
\usepackage{mathabx}
\usepackage[sort&compress,numbers]{natbib}
\usepackage{authblk}

\definecolor{Orange}{rgb}{1,0.64,0}
\definecolor{lgray}{rgb}{0.9,0.9,0.9}

\makeindex

\begin{document}

\title{Multifractal Characterization of Protein Contact Networks}

\author[1]{Enrico Maiorino\thanks{enrico.maiorino@uniroma1.it}\thanks{Corresponding author}}
\author[2]{Lorenzo Livi\thanks{llivi@scs.ryerson.ca}}
\author[3]{Alessandro Giuliani\thanks{alessandro.giuliani@iss.it}}
\author[2]{Alireza Sadeghian\thanks{asadeghi@ryerson.ca}}
\author[1]{Antonello Rizzi\thanks{antonello.rizzi@uniroma1.it}}
\affil[1]{Dept. of Information Engineering, Electronics, and Telecommunications, SAPIENZA University of Rome, Via Eudossiana 18, 00184 Rome, Italy}
\affil[2]{Dept. of Computer Science, Ryerson University, 350 Victoria Street, Toronto, ON M5B 2K3, Canada}
\affil[3]{Dept. of Environment and Health, Istituto Superiore di Sanit\`{a}, Viale Regina Elena 299, 00161 Rome, Italy}
\renewcommand\Authands{, and }
\providecommand{\keywords}[1]{\textbf{\textit{Index terms---}} #1}

\maketitle

\begin{abstract}
The multifractal detrended fluctuation analysis of time series is able to reveal the presence of long-range correlations and, at the same time, to characterize the self-similarity of the series.
The rich information derivable from the characteristic exponents and the multifractal spectrum can be further analyzed to discover important insights about the underlying dynamical process.
In this paper, we employ multifractal analysis techniques in the study of protein contact networks.
To this end, initially a network is mapped to three different time series, each of which is generated by a stationary unbiased random walk. To capture the peculiarities of the networks at different levels, we accordingly consider three observables at each vertex: the degree, the clustering coefficient, and the closeness centrality.
To compare the results with suitable references, we consider also instances of three well-known network models and two typical time series with pure monofractal and multifractal properties.
The first result of notable interest is that time series associated to proteins contact networks exhibit long-range correlations (strong persistence), which are consistent with signals in-between the typical monofractal and multifractal behavior.
Successively, a suitable embedding of the multifractal spectra allows to focus on ensemble properties, which in turn gives us the possibility to make further observations regarding the considered networks. In particular, we highlight the different role that small and large fluctuations of the considered observables play in the characterization of the network topology.\\
\keywords{Multifractal analysis; Complex biological networks; Time series analysis; Random walk.}
\end{abstract}

\section{Introduction}
\label{sec:intro}

Beside the prophecy made by \citet{weaver1991science}, identifying in the study of ``organized complexity'' (i.e., systems made by many interconnected elements) as the new theoretical frontier of Sciences, the Dutch electrical engineer Bernard Tellegen, already in 1952 demonstrated the existence of regularities and laws dependent only on the wiring architecture of the studied systems \cite{mikulecky2001network,tellegen1952general}.
These regularities cause systems made by elements obeying different physical laws (e.g. electrical circuits, molecules, metabolic networks) to display a very similar mesoscopic behavior. These ``network laws'' are still largely unknown and there is a huge room of investigation for catching these principles.
This is the main focus of the emerging research field of complex networks analysis \cite{wei2014new,havlin2010,dorogovtsev2008critical,xiao2008symmetry,lacasa2013correlation,lee2006statistical,costa2007characterization}.

A well-known approach to characterize complex networks consists in analysing the fractal properties of their topology (i.e., scaling properties), which is usually implemented by means of the box counting method \cite{song2005self}.
Recently, new methods \cite{dan2012multifractal,li2014fractal,furuya2011multifractality} have been proposed to generalize this analysis to the multifractal setting \cite{harte2010multifractals}.
A network (or equivalently, a graph) is a very general mathematical construct having the power of unifying apparently different objects like correlation matrices, recurrence plots, adjacency matrices. The only necessary property to call something a graph is the existence of a wiring scheme that, for any pair of constituting elements, returns a binary information (or in certain cases, the strength) about the existence of a link between the elements. 
The definition of this wiring scheme allows to transfer virtually any problem to the graph domain, granting the possibility to take advantage of all the tools coming from classical graph theory and the more recently developed complex networks analysis.
In particular, when studying time series it is possible to generate a network (in the form of recurrence plots, correlation matrices, or other mapping methods) that preserves their temporal correlation structure in the form of topological invariants \cite{donner2011,marwan2007recurrence,campanharo2011duality}.
Nonetheless, the time series / network ``isomorphism'' works in both directions. The exploration of large networks by means of a random walker, taking at each vertex different directions along the graph topology according to some probabilistic criteria \cite{bonaventura2014characteristic}, has the potential to offer a dynamical perspective of the studied network.
However, while the former approach has been deeply investigated \cite{donner2011}, less attention has been devoted to the latter in the case of (multi)fractal characterization, apart from few recent exceptions \cite{nicosia2013characteristic,weng2014time,zhou2014fractal}.

In this paper we exploit the Multifractal Detrended Fluctuation Analysis (MFDFA) \cite{kantelhardt2002multifractal,PhysRevE.74.016103,bashan2008comparison}, a generalization of the Detrended Fluctuation Analysis (DFA) \cite{peng1995quantification}, to study time series obtained from complex networks via stationary unbiased random walks (RW).
The MFDFA builds upon a generalization of the so-called Hurst exponent as a detector of long-range correlations \cite{serinaldi2010use,barunik2010hurst}.
At the basis of Hurst exponent is the idea of characterizing time series in terms of their degree of \textit{persistence}: roughly speaking, a series is long-range correlated (persistent) if the underlying process has memory of the past states, a property that is firstly noticeable as a heavy-tail in the corresponding autocorrelation function.
Brownian motion corresponds to Hurst exponent equal to 0.5 and it is considered as the baseline uncorrelated process.
Series with Hurst exponent greater than 0.5 are considered as persistent; series with Hurst exponent smaller than 0.5 are anti-persistent (consecutive values tend to be very different).
Additionally, if the value of this exponent does not vary significantly with the magnitude of fluctuations, then the time series is considered monofractal and it can be consistently analysed via DFA; in the opposite case, it is multifractal and the MFDFA is a more suitable choice.
If the studied time series corresponds to a sequence of discrete observables attached to the vertices of a network and the ordering is determined by the subsequent encounters of a random walker exploring the graph, then its persistence / antipersistence property can be translated into the assortative / disassortative character of the graph with respect to said observables.
An assortative graph \cite{newman2002assortative} is a graph in which vertices with similar properties (typically the degree is used, but in theory any property of the vertex can be taken into account) tend to be in contact more frequently than what expected by chance, while a disassortative graph has the opposite feature.
Studying a complex network by the action of a random walker producing a collection of time series of encounters with vertices has an advantage with respect to the simple computation of the static assortative indexes of the graph.
Indeed, the walker trajectories offer also a sampling of the paths distribution in the graph. This distribution is affected by the whole set of mutual relations of vertices at different scales, which are not fully appreciable by a single static snapshot of the network by means of classical network invariants.
In the same manner, we are able to gain an insight on the different scaling of the autocorrelation function and hence on the distribution of the corresponding observable across different locations and scales of the network.

In this study, we primarily focus on the protein contact networks (PCN) elaborated from the E. coli \cite{ecoli_graph__arxiv}. The networks consist of amino acid residues put in contact according to the 3D protein structure \cite{doi:10.1021/cr3002356}. We compare the properties of PCN with those of different known network and time series models.
In particular, we bias the study on their analogies and differences with a model of synthetic protein contact networks (PCN-S) as theorysed by \citet{bartoli2007effect}; PCN-S consist in coiled coils of polymers in which the probability of contact is a decreasing function of the distance between residues along the chain.
From our study, PCN clearly emerge as a particular class of networks whose time series exhibit properties in between mono and multi fractal signals.
The synthetic (simulated coiled coils) proteins, i.e., PCN-S, preserve sufficiently well the multifractality (thus the presence of different scaling relations at different scales or positions), while they seem to lose much of the strong persistence typical of PCN time series. Comparing the PCN with their synthetic counterparts allowed us to highlight a different level of order of the former not shown in the latter due to the pure synthetic polymer folding in space (i.e. no excluded volume contraints). Indeed, the additional freedom given by pure folding causes the loss of many peculiar properties of proteins, like allosteric effect (i.e. the global configurational change of the system upon ligand binding at a specific site) and rapid folding \cite{hilser2012structural,tsai2009protein}. Notably, the particular wiring of protein contact networks encompasses two different architectures fused in the same graph: a short-range contact pattern linking aminoacid residues within the same module, and a long-range contact pattern linking aminoacid residues very distant on the chain structure and pertaining to different modules. The long-range contacts are much less in number than short-range ones but allow for the rapid spreading of information across the structure in order to make the allosteric effect and rapid folding possible.  
We successively focus on the embedding of multifractal spectra (MFS) domain and codomain in a vector space to offer further insights on the sensitivity of the coefficients describing the MFS with respect to fluctuation magnitudes (i.e., the small and large fluctuations of the time series, expressed by the $q$th-order moments of the variances).
The discriminating capabilities of different orders of fluctuations of the time series are highlighted by a principal component analysis (PCA) of the embedded MFS, which reduces the initial high-dimensional MFS profiles to four principal components (PC).
Interestingly, the result is that short and medium range observables (respectively, vertex degree and clustering coefficient) are influenced by the network class mainly in their large fluctuations, while the closeness centrality is sensitive to the topology at all fluctuation scales, given its global nature.
PCA of the MFS is hence proposed as a convenient post-processing method when the qualitative interpretation of the results is a primary objective in a multifractal analysis.

The remainder of this paper is structured as follows.
In Section \ref{sec:mfdfa} we provide the necessary technical background on the MFDFA.
In Section \ref{sec:data} we describe the data that we considered in this study.
Section \ref{sec:results} presents the results, which are organized in three subsections.
In Section \ref{sec:persistence} we study the persistence properties of the time series; Section \ref{sec:multifractal} offers a study on the MFS meant to provide a first insight about the multifractality of the time series; finally in Section \ref{sec:embedding} we discuss and interpret the statistical properties of the embedding space derived by means of PCA of the MFS.
Section \ref{sec:conclusions} provides the conclusions and offers some future directions.

\section{Multi-Fractal Detrended Fluctuation Analysis}
\label{sec:mfdfa}

It is known that many processes in Nature and society present long-term memory, manifested in primis as heavy tails in the autocorrelation function of the considered observables. This phenomenon, referred to as the \textit{persistence} of a process, can be characterized by the value of the Hurst exponent $H$, introduced in 1951 by the British hydrologist Harold Edwin Hurst \cite{hurst1951long}. The exponent normally assumes values in the range $[0,1]$ and is traditionally calculated with the R/S analysis, as shown in \cite{serinaldi2010use}. When the process corresponds to uncorrelated noise (e.g. Brownian motion) then the value of $H$ is 0.5, whereas if the process is persistent (correlated) or antipersistent (anticorrelated) it will be respectively greater than and less than 0.5. However, conventional methods employed to analyze the long-range correlation properties of a time series (e.g., spectral analysis, Hurst analysis \cite{serinaldi2010use,barunik2010hurst}) reveal to be 
misleading when said time series is non-stationary.
In fact, in many cases it is important to distinguish fluctuations caused by trending behaviors of data at all time scales -- which in this context can be regarded as noise -- from the intrinsic fluctuations characterizing the dynamical process generating the time series.
One of the methods usually employed for this purpose is the Detrended Fluctuation Analysis (DFA), which has shown to be successful in a broad range of situations \cite{peng1995quantification}.
The DFA has been generalized in the so-called Multifractal Detrended Fluctuation Analysis (MFDFA) \cite{kantelhardt2002multifractal,PhysRevE.74.016103,bashan2008comparison,PhysRevLett.62.1327}, which accounts for multifractal scaling, that is, different correlation behaviors on different portions of data, which are thus identified by different sets of scaling exponents.
Among the many applications of MFDFA, it is possible to cite the analysis of human EEG \cite{zorick2013multifractal}, solar magnetograms \cite{makarenko2012multifractal}, human behavioral response \cite{ihlen2013multifractal}, hippocampus signals \cite{fetterhoff2014multifractal}, seismic series \cite{telesca2006measuring,telesca2005multifractal}, medical imaging \cite{lopes2009fractal}, financial markets \cite{schmitt1999multifractal,barunik2012understanding}, and written texts \cite{PhysRevE.86.031108}.

The MFDFA procedure is described thoroughly in \cite{kantelhardt2002multifractal} and it is reported briefly in the following. The method can be summarized in five steps, three of which are identical to the DFA version.
Given a time series $x_k$ of length $N$ with compact support, the MFDFA steps are:
\begin{itemize}
\item{\textit{Step 1} : Compute $Y(i)$ as the cumulative sum (profile) of the series $x_k$:
\begin{equation}
Y(i) \equiv \sum_{k=1}^i \left[ x_k - \langle x \rangle \right], \;\; i = 1, \dotsc, N.
\end{equation}
}
\item{\textit{Step 2} :
Divide $Y(i)$ in $N_s \equiv \text{int}(N/s)$ non-overlapping segments of equal
length $s$. Since the series length $N$ may not be a multiple of $s$, the last segment is likely to be shorter, so this operation is repeated in reverse order by starting from the opposite end of the series, thus obtaining a total of $2N_s$ segments.}
\item{\textit{Step 3} : Execute the local detrending operation by a suitable polynomial fitting on
each of the $2N_s$ segments. Then determine the variance,
\begin{equation}
F^2(\nu,s) \equiv \frac{1}{s} \sum_{i=1}^s \bigg\{ Y[(\nu-1)s+i] - y_\nu(i) \bigg\}^2,
\end{equation}
for each segment $\nu = 1,\dotsc,N_s$ and
\begin{equation}
F^2(\nu,s) \equiv \sum_{i=1}^s \bigg\{ Y[N-(\nu - N_s)s+i] - y_\nu(i)\bigg\}^2
\end{equation}
for $\nu = N_s +1,\dotsc, 2N_s$, where $y_\nu(i)$ is the fitted polynomial in segment $\nu$. The order $m$ of the fitting polynomial, $y_\nu(i)$, determines the capability of the (MF-)DFA in eliminating trends in the series, thus it has to be tuned according to the expected maximum trending order of the time series.
}
\item{\textit{Step 4} : Compute the $q$th-order average of the variance over all segments,
\begin{equation}
\label{eq:Fq}
F_q(s) \equiv \bigg\{ \frac{1}{2N_s} \sum_{\nu=1}^{2N_s} \left[ F^2(\nu,s)\right]^{q/2} \bigg\}^{1/q},
\end{equation}
with $q \in \mathbb{R}$. The $q$-dependence of the fluctuations function $F_q(s)$ serves the purpose of highlighting the contribute of fluctuations at different magnitude orders.
For $q > 0$ only the larger fluctuations contribute mostly to the average in Eq.~\ref{eq:Fq}; conversely, for $q < 0$ the magnitude of the smaller fluctuations is enhanced. For $q = 2$ the standard DFA procedure is obtained. The case $q = 0$ cannot be computed with the averaging form in Eq.~\ref{eq:Fq} and so a logarithmic form has to be employed,
\begin{equation}
F_0(s) = \exp \bigg\{ \frac{1}{2N_s} \sum_{\nu=1}^{2N_s} \ln \left[F^2(\nu,s)\right] \bigg\}.
\end{equation}
The steps 2 to 4 have to be repeated for different time scales $s$, where all values of $s$ have to be chosen such that $s \geq m+2$ to allow for a meaningful fitting of data. It is also convenient to avoid scales $s > N/4$ because of the statistical unreliability of such small numbers $N_s$ of segments considered.
}
\item{\textit{Step 5} : Determine the scaling behaviour of the fluctuation functions by analyzing log-log plots of $F_q(s)$ versus $s$ for each value of $q$. If the series $x_i$ is long-range power-law correlated, $F_q(s)$ is approximated (for large values of $s$) by the form
\begin{equation}
\label{eq:Fqshq}
F_q(s) \sim s^{h(q)}.
\end{equation}} 
\end{itemize}

The exponent $h(q)$ is the generalized Hurst exponent; for $q=2$ and stationary time series, $h(q)$ reduces to the standard Hurst exponent, $H$. When the considered time series is monofractal, i.e., it shows a uniform scaling over all magnitude scales of the fluctuations, $h(q)$ is independent of $q$. On the contrary, when small fluctuations are different with respect to the large ones, the dependency of $h(q)$ on $q$ becomes apparent and the series can be considered multifractal.

Starting from Eq.~\ref{eq:Fq} and using Eq.~\ref{eq:Fqshq}, it is straightforward to obtain
\begin{equation}
\sum_{\nu=1}^{N/s} [ F(\nu,s)]^q \sim s^{qh(q) - 1},
\end{equation}
where, for simplicity, it has been assumed that the length $N$ of the series is a multiple of the scale $s$, such that $N_s = N/s$.
The exponent
\begin{equation}
\label{eq:tauq}
\tau(q) = qh(q) - 1
\end{equation}
corresponds to the MFDFA generalization of the multifractal mass exponent. In case of positive stationary and normalized time series, $\tau(q)$ corresponds to the scaling exponent of the $q$-order partition function $Z_q(s)$.
Another function that characterizes the multifractality of a series is the singularity spectrum, $D(\alpha)$, which is obtained via the Legendre transform of $\tau(q)$,
\begin{equation}
\label{eq:mutifractal_spectrum}
D(\alpha) = q\alpha - \tau(q),
\end{equation}
where $\alpha$ is equal to the derivative $\tau'(q)$ and corresponds to the H\"older exponent (also called singularity exponent). Using Eq.~\ref{eq:tauq} it is possible to directly relate $\alpha$ and $D(\alpha)$ to $h(q)$, obtaining:
\begin{equation}
\alpha = h(q) + qh'(q) \;\; \text{and} \;\; D(\alpha) = q[\alpha - h(q)] + 1.
\end{equation}

The singularity spectrum in Eq. \ref{eq:mutifractal_spectrum} -- also called multifractal spectrum (MFS) -- allows to infer important information regarding the ``degree of multifractality'' and the specific sensitivity of the time series to fluctuations of different magnitudes.
In fact, the width of the support of $D(\cdot)$ is an important quantitative indicator of the multifractal character of the series (the larger, the more fractal a series is).
Also the codomain of $D(\cdot)$ encodes useful information, since it corresponds to the dimension of the subset of the times series domain which is characterized by the singularity exponent $\alpha$.

\section{The Considered Data}
\label{sec:data}

The biological networks analysed in this work are partially linked to those analysed in our previous works \cite{mixbionets1__arxiv,ecoli_graph_complexity_arxiv}. We consider 400 E. coli protein contact networks (PCN) as the main object of study and we compare them to several models.
To this end, we generated 400 synthetic polymers (PCN-S) by employing the generation method presented by \citet{bartoli2007effect}, and by setting appropriate parameters in order to resemble the basic properties of each of the above PCN (i.e., the graph size).
More precisely, each E. coli protein {\tt JWxxxx} is juxtaposed with its synthetic counterpart, {\tt JWxxxx\,\_\,SYNTH}, having equal number of vertices and edges -- the four-digit number {\tt xxxx} stands for its unique identifier. 
In addition, we consider 10 Erd\H{o}s-R\'{e}nyi networks (ER) and 10 scale-free networks generated using the Barab\'{a}si-Albert (BA) model, varying the number of vertices between 300 and 1200. The former are generated setting $p=\log(n)/n$ while for the latter we used a six-degree attachment scheme.
To allow the processing of such networks via the MFDFA procedure, we generate time series by means of stationary unbiased RWs, where at each step an observable is measured from the current vertex. Considering the size of the networks at hand, the RW length has been fixed at $10^5$ time instants; this length assures the coverage of all vertices for a statistically significant number of times and it is consistent with the recommendations in Ref. \cite{nicosia2013characteristic}.
We associate to each network three time series generated within the same RW. The first series considers vertex degree (VD) as observable; the second one the vertex clustering coefficient (VCL); the third one the vertex closeness centrality (VCL).
Those three observables account for, respectively, the short, medium, and long range information of the network from the point of view of a vertex.

The dataset is also composed by six classes of time series that act as probes, which are obtained directly from their generative models.
The herein considered time series are obtained from three fractional Brownian motion (FBM) processes with increasing Hurst coefficients, and three multifractal binomial cascades (BC), characterized by increasing MFS widths. FBMs have coefficients $H = 0.25,0.5,0.75$ and represent the poles of monofractality with increasing persistence. For each fixed value of $H$, we generated ten different time series (for a total of 30 FBMs) to account for the statistical variability. 
On the other hand, BCs are deterministic multiplicative processes, which are generated with the partition coefficient $a = 0.6,0.7,0.8$. These series are inherently multifractal, although they possess different persistence levels. Notice that in this case there is no point in generating more than one instance of the BC processes for each value of $a$, since the process is deterministic; so only three BC time series are generated.

\section{Experimental Results}
\label{sec:results}

In this section we discuss the experimental results obtained by analyzing all considered time series. The results are organized in three main sections. In Sec. \ref{sec:persistence} we show the analysis of the persistence properties of the time series; Sec. \ref{sec:multifractal} provides the setting of the procedure employed to obtain the MFS coefficients, offering also an initial discussion on the multifractal properties of the considered time series; finally in Sec. \ref{sec:embedding} we discuss the embedding of the MFS in a suitable vector space.

\subsection{Analysis of persistence properties}
\label{sec:persistence}

\begin{figure}[ht!]
\centering

\subfigure[Time series of VD.]{
\includegraphics[viewport=70 0 340 190,scale=0.55,keepaspectratio=true]{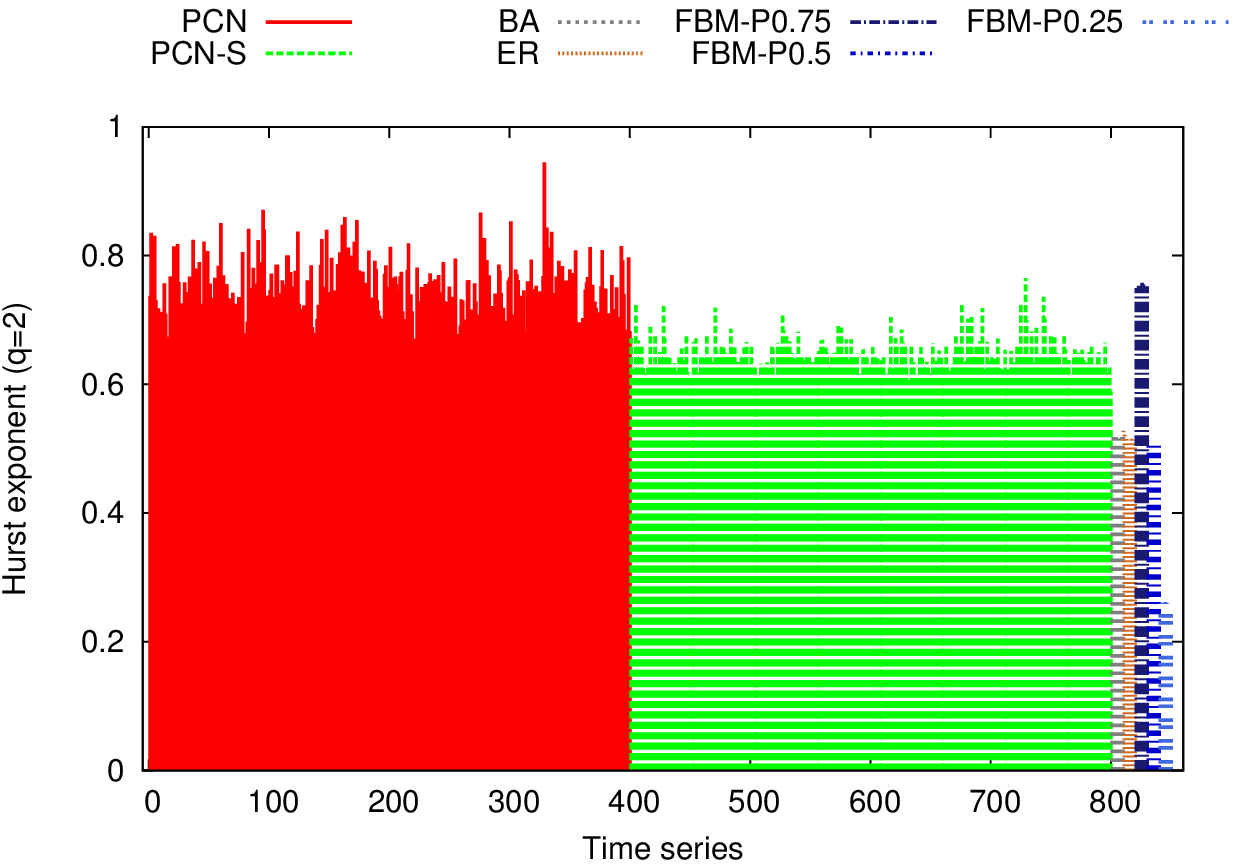}
\label{fig:persistence_deg}}
~
\subfigure[Time series of VCL.]{
\includegraphics[viewport=20 0 300 190,scale=0.55,keepaspectratio=true]{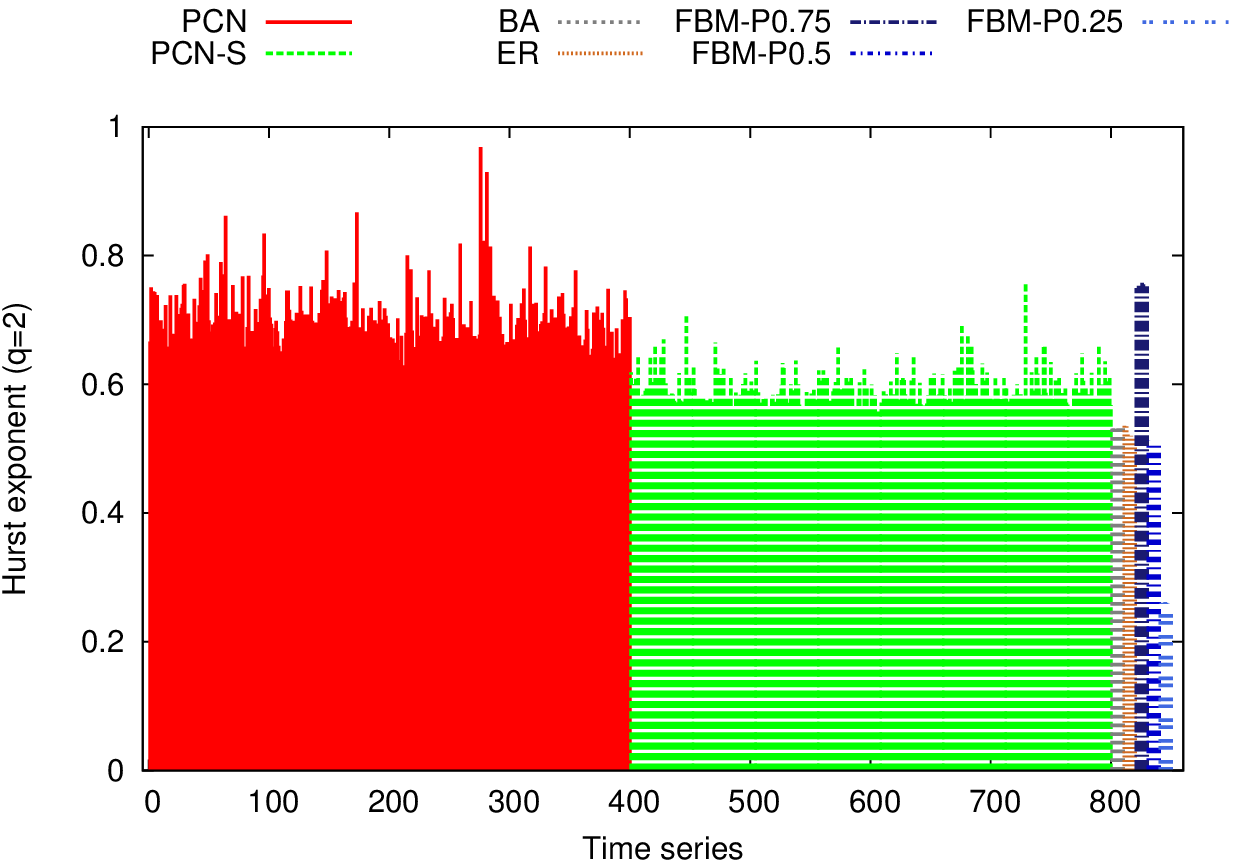}
\label{fig:persistence_cl}}

\subfigure[Time series of VCC.]{
\includegraphics[viewport=75 0 340 240,scale=0.55,keepaspectratio=true]{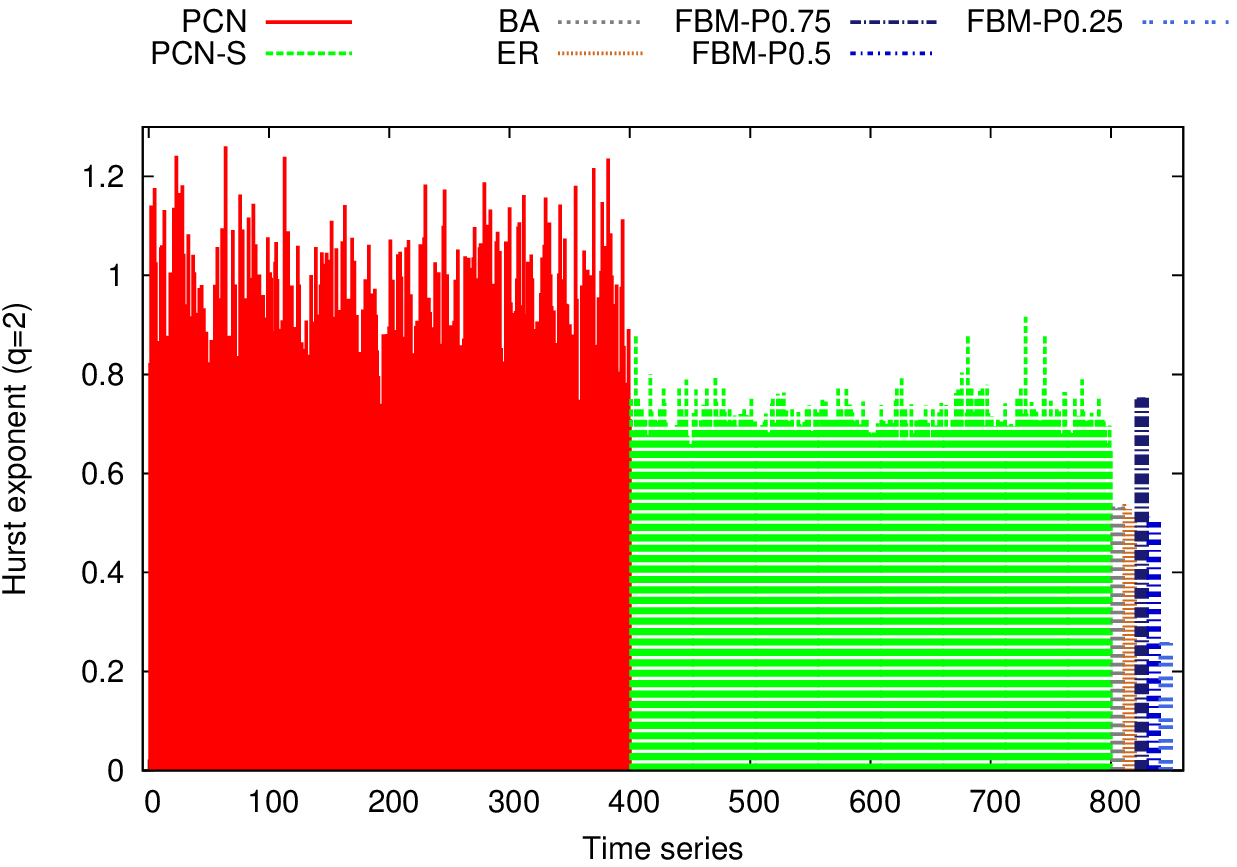}
\label{fig:persistence_cc}}\hspace{0.1cm}
~
\subfigure[Autocorr. for ``JW0058''.]{
\includegraphics[viewport=20 0 300 190,scale=0.53,keepaspectratio=true]{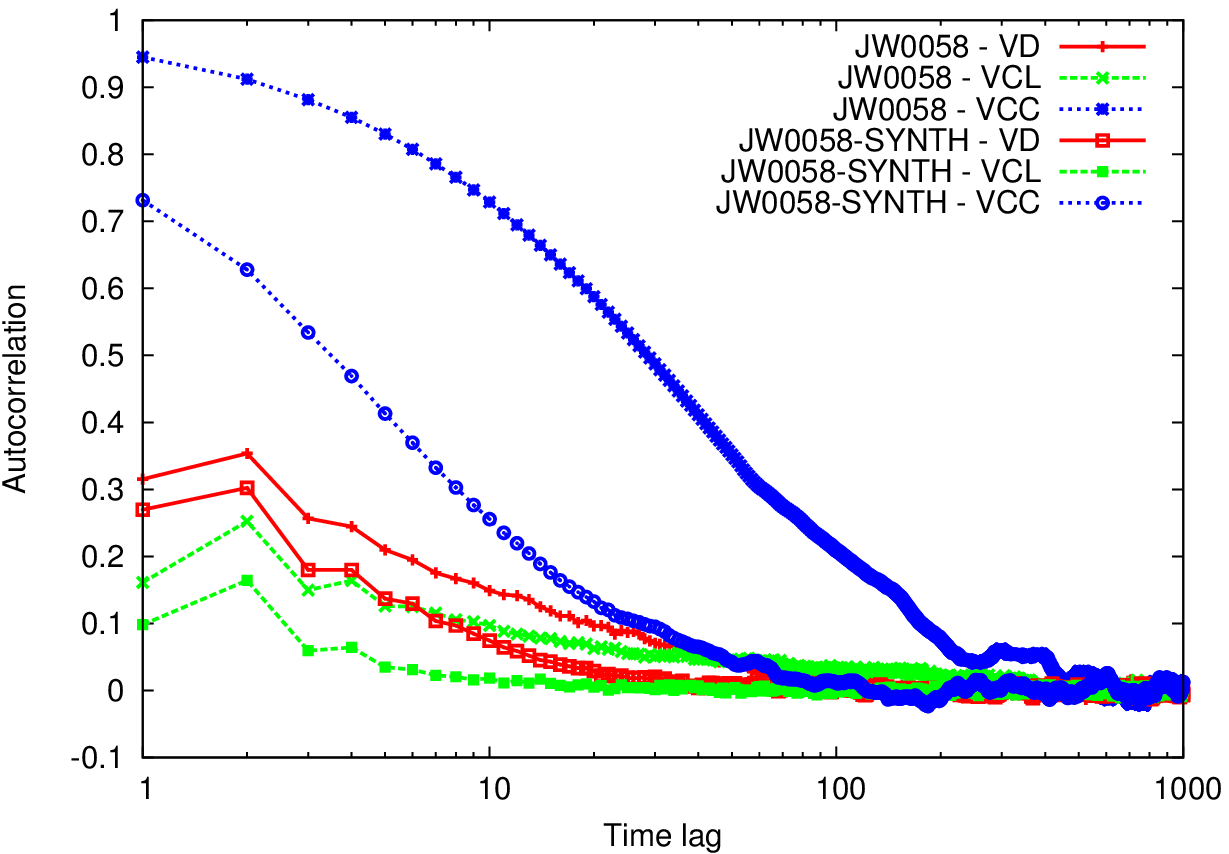}
\label{fig:autocorrelation}}

\subfigure[Time series for ``JW0058''.]{
\includegraphics[viewport=0 0 341 177,scale=0.50,keepaspectratio=true]{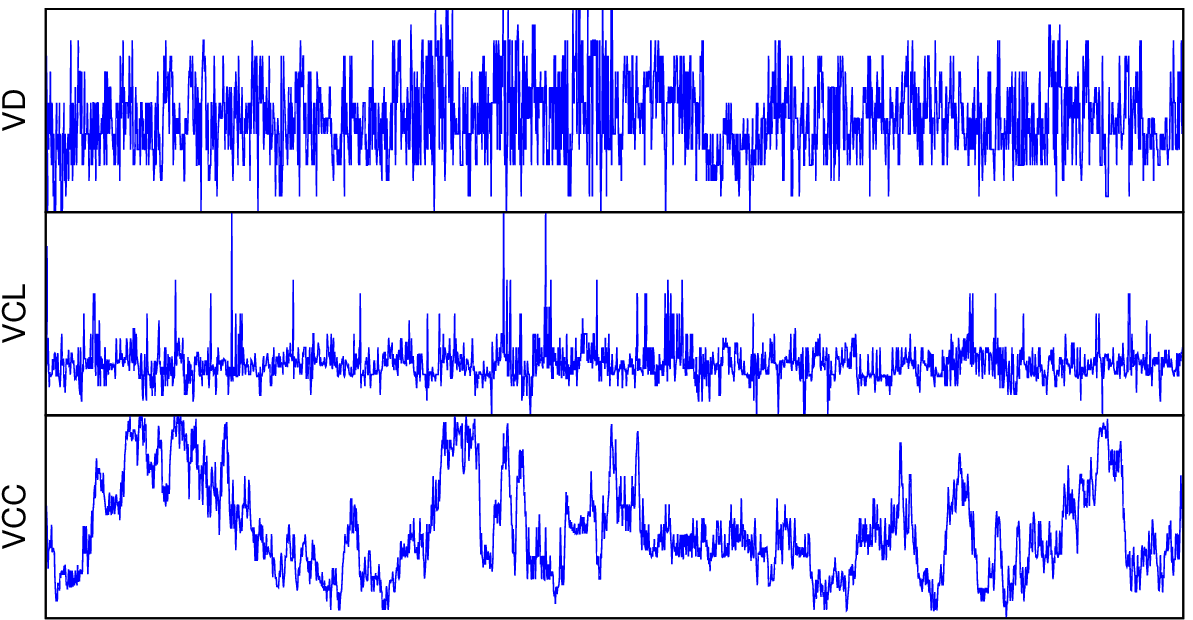}
\label{fig:tsjw0058}}
~
\subfigure[Time series for ``JW0058\_SYNTH''.]{
\includegraphics[viewport=0 0 341 177,scale=0.50,keepaspectratio=true]{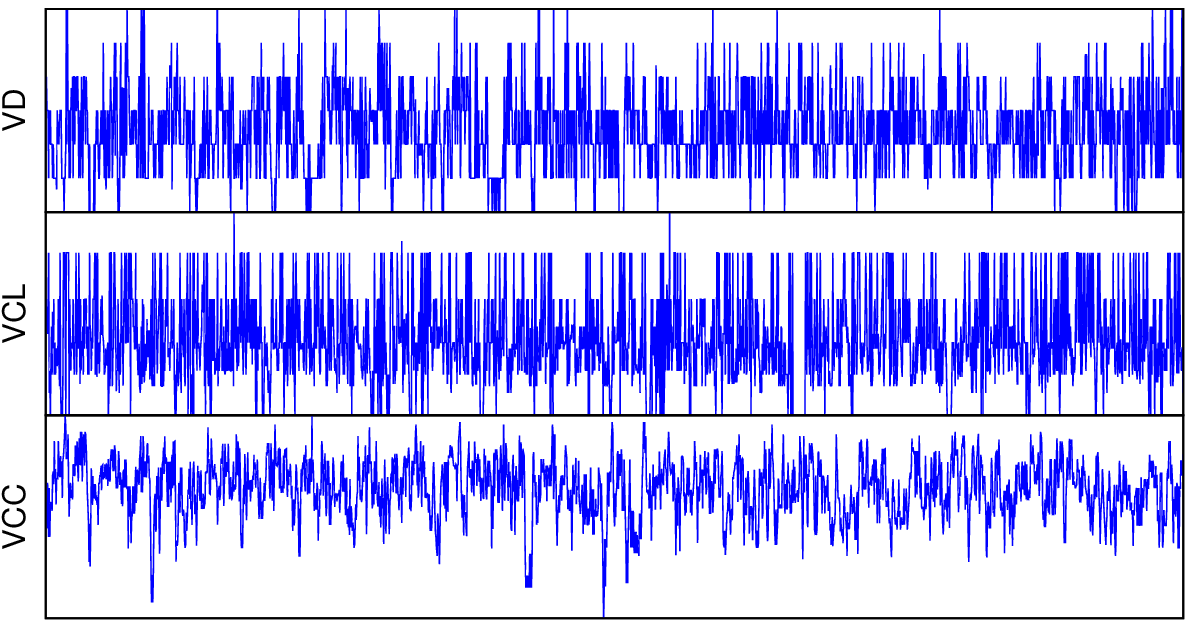}
\label{fig:tsjw0058s}}

\caption{Persistence of the series measured through the Hurst exponent for VD \subref{fig:persistence_deg}, VCL \subref{fig:persistence_cl}, and VCC \subref{fig:persistence_cc}. The PCN (red bands) show significant persistence for all the three observables. \subref{fig:autocorrelation} Sample autocorrelation function for the protein JW0058 and the corresponding synthetic polymer JW0058\_SYNTH. \subref{fig:tsjw0058} and \subref{fig:tsjw0058s} Sample time series. The higher persistence of the natural protein with respect to the synthetic analogue is particularly evident in the VCC series.}
\label{fig:persistence_level}
\end{figure}

The first property that we analyze is the Hurst coefficient that, as described above, quantifies the persistence of the time series.
In Figs.~\ref{fig:persistence_deg}, \ref{fig:persistence_cl}, and \ref{fig:persistence_cc} are shown the values of $H$ measured on each time series of the PCN, PCN-S, BA, and ER, for each of the three observables VD, VCL, and VCC, respectively, along with the Hurst exponents proper of the three classes of FBMs. Notice that, since FBM time series are not obtained as different observables yielded by a RW on a network, their Hurst exponents have been just replicated across the three figures.

As expected, BA and ER networks produce RWs consistent with an uncorrelated Brownian motion (i.e., $H=0.5$), since basically they are the result of an uncorrelated degree distribution.
Interestingly, from the persistence levels shown in Fig.~\ref{fig:persistence_level}, it is possible to observe that PCN (red bands) induce time series with strong persistence, regardless of the particular observable. It is also evident from Fig.~\ref{fig:persistence_level} that also synthetic polymers (green bands), similarly to the PCN, show positively correlated behaviours, even if they do not seem to capture this characteristic persistence in a satisfying manner. It is also important to mention that, when plotting the Hurst exponents of PCN-S as a function of Hurst exponents of their corresponding PCN (data not shown for brevity), no trending has been observed. Indeed, these two quantities are not proportional and thus PCN-S instances cannot be considered just as less-persistent versions of their corresponding PCN.    
These results can be exploited to gain a more insightful view on the intrinsic characteristics of the PCN class, by relating the properties of the RWs to the topological properties of the corresponding graphs. In particular, the time series of VD show positive correlations, which in turn imply degree assortativity. This result is in agreement with the claims of \citet{bode2007network} and references therein, although we reached the same result by exploiting a different technique -- usually the degree assortativity is investigated through the method proposed by \citet{newman2002assortative}.
It is worth pointing out that, since PCN are known to be fractal networks (embedded into a three-dimensional space) \cite{granek2011proteins,ecoli_graph_complexity_arxiv}, the herein observed degree assortativity is not in agreement with the theoretical hypothesis of \citet{song2006origins}, which requires the degree distribution of fractal networks to be disassortative.

\begin{figure}[ht!]
\centering

\subfigure[VCL vs VD.]{
\includegraphics[viewport=0 0 347 248,scale=0.60,keepaspectratio=true]{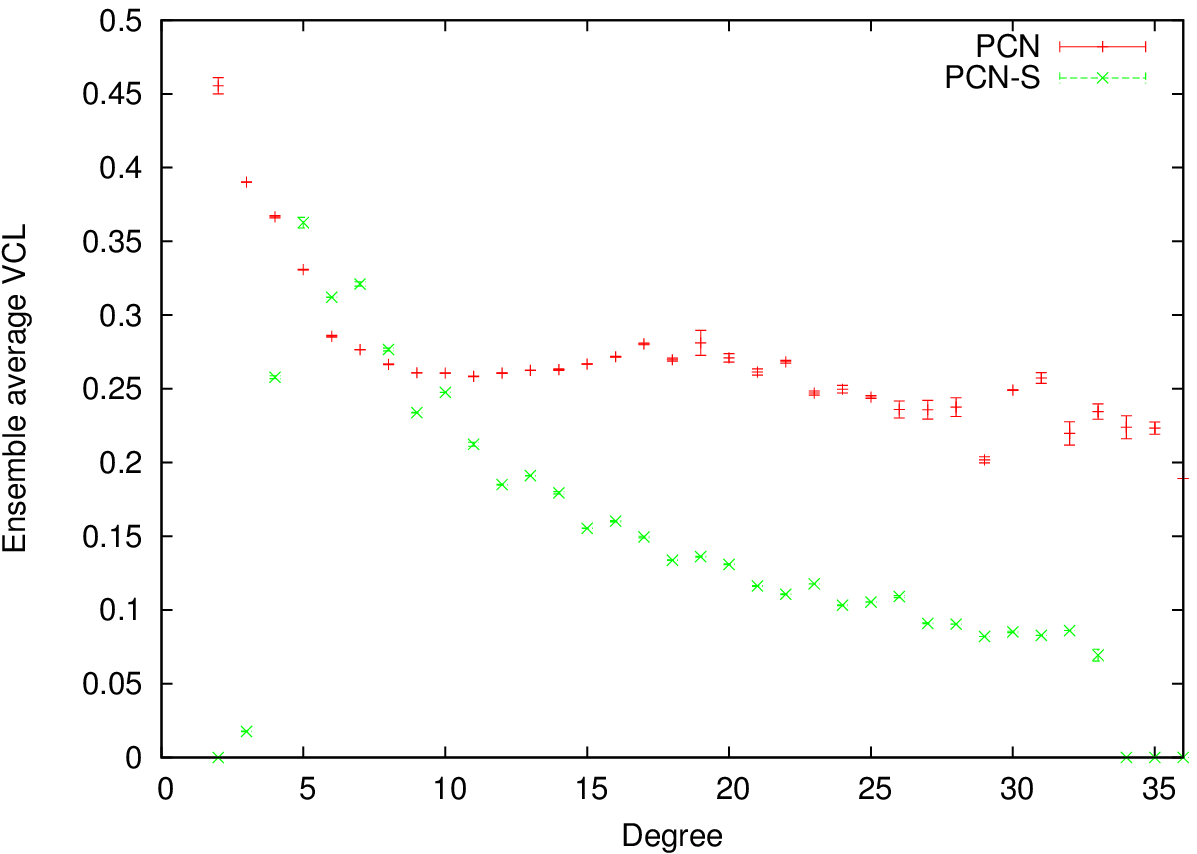}
\label{fig:degavgcl}}
~
\subfigure[VCC vs VD.]{
\includegraphics[viewport=0 0 347 248,scale=0.60,keepaspectratio=true]{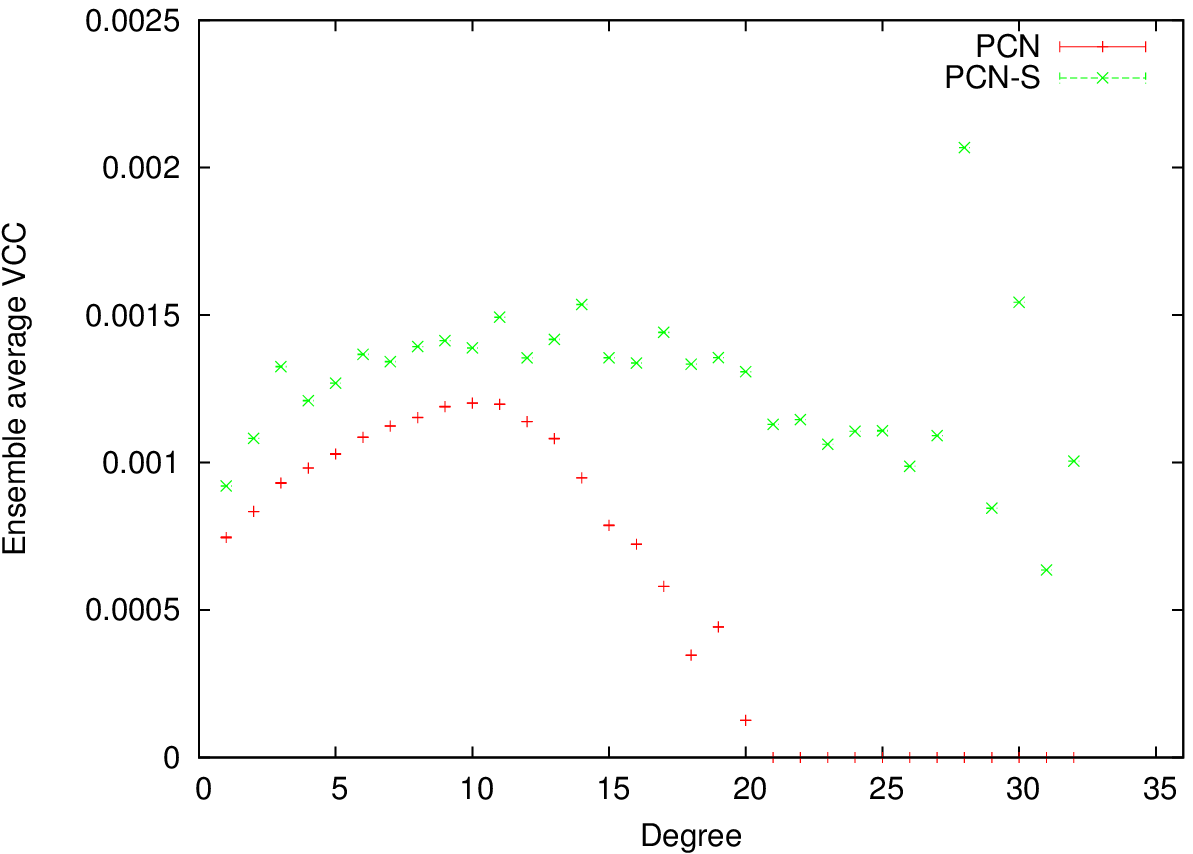}
\label{fig:degavgcc}}

\caption{Ensemble average VCL \subref{fig:degavgcl} and VCC \subref{fig:degavgcc} as a function of the degree for all proteins (red points) and synthetic counterpart (green points). Relation with respect to the degree shows significant difference among the real proteins and the synthetic polymers.}
\label{fig:avg_vs_deg}
\end{figure}

The high persistence of the clustering coefficient observed in the PCN is slightly more tricky to interpret in terms of topological properties.
Roughly speaking, the VCL of a vertex is proportional to the local connectivity of the subgraph formed by the vertex and its closest neighbors with respect to the whole graph.
It is known that PCN show a high degree of global modularity (see \cite{mixbionets1__arxiv,doi:10.1021/cr3002356}). Therefore, the persistence of the clustering coefficient can be interpreted as the tendency of vertices in the same module to be connected rather uniformly with the presence of medium-to-small hubs -- PCN do not have large hubs \cite{bode2007network,doi:10.1021/cr3002356}.
Another way to explain this property is to directly relate VCL to the persistence of VD time series.
To this end, Fig.~\ref{fig:degavgcl} shows the relation of the degree-dependent average VCL over the possible VD. Here we considered the whole PCN and PCN-S ensembles. The error bars (which are usually smaller than the marker) represent the standard deviation over the entire ensemble.
As it is possible to observe, while the two trends are substantially different, the standard deviation is very small in both cases for most values of the degree. This fact, along with the aforementioned degree assortativity, suggests a possible explanation for the persistence displayed by the VCL series of both PCN and PCN-S.
It is worth noting that, for PCN, the clustering coefficient remains high when increasing the degree, which can be interpreted as a sign of high global modularity.

While VD and VCL show similar characteristics, the behaviour of the VCC observable is considerably different. By looking at the plot in Fig.~\ref{fig:persistence_cc}, it is possible to observe that the Hurst coefficients of PCN are comparable and occasionally greater than one -- i.e., the corresponding time series are non-stationary. This might be considered as a symptom of different distributions of typical paths within the PCN.
This conjecture would be in line with the observation of \citet{yan2014construction}, where it is hypothesized that there are two characteristic distributions of paths within PCN, intra-module and inter-module paths, which is also a consequence of the PCN's high degree of modularity mentioned before.
On the other hand, PCN-S do not share this feature, confirming an intrinsically different configuration of the network topology at a global scale.
As for the VCL observable, the VCC persistence can be related to the VD persistence by inspecting Fig.~\ref{fig:degavgcc}.
By comparing the two trends, it can also be observed that the PCN-S have a broader distribution over the possible degree values, as a consequence of being small-world networks \cite{bartoli2007effect}, while PCN are neither small-world nor scale-free \cite{mixbionets1__arxiv,doi:10.1021/cr3002356}.
In Fig.~\ref{fig:autocorrelation} it is shown the autocorrelation function of the three time series for one randomly chosen protein and its synthetic counterpart.
First, it is worth noting that long-range correlations appear here in the form of heavy tails.
Additionally, the VCC autocorrelation function denotes a much heavier tail with respect to the other observables, which is justified by the higher persistence (see Fig. \ref{fig:persistence_cc}).
To conclude, in Fig.~\ref{fig:tsjw0058} and \ref{fig:tsjw0058s} are shown two excerpts of the time series generated by the same protein and its synthetic model for each observable. By visually comparing the two plots, in particular for the observables VCL and VCC, it is possible to note that the two networks generate RWs that are significantly different; the higher persistency of the PCN observables is also visually recognizable. 

From these results it is clear that the PCN-S network models present significant discrepancies from their real counterparts, while still being distinguishable from other network models. These differences will be further analyzed in the following subsections.

\subsection{Analysis of multifractal properties}
\label{sec:multifractal}

After having calculated the persistence properties of the considered time series, we can now proceed to evaluate their degree of multifractality.
For each of the time series presented in Sec. \ref{sec:data}, we perform the MFDFA procedure exposed in Sec. \ref{sec:mfdfa} by executing the Matlab$^\circledR$ routine {\tt MFDFA1()}, written by Ihlen and described in detail in Ref. \cite{ihlen2012introduction}.
The input of the routine is the time series to analyze, a vector of the considered time scales (corresponding to the set of increasing length scales $s$ described in Sec. \ref{sec:data}), the range of $q$-orders to be considered for the analysis, and finally the polynomial order, $m$, for the detrending.
For the analysis of all time series, we used the following setting:
\begin{itemize}
\item{the time scales $s \in \{16,32,64,128,256,512,1024\}$;}
\item{the orders $q \in \{ -5, -4.8, -4.6, \dotsc, +4.8, +5\}$ for a total of 51 values;}
\item{the detrending order $m = 2$.}
\end{itemize}

The output produced by the routine, for all values of $q$, is the collection of (generalized) Hurst coefficients $H(q)$, mass exponents $\tau(q)$, singularity exponents $\alpha(q)$, dimension coefficients $D(\alpha(q))$, and scaling function $F(q)$.
Please note that since $D(\alpha(q))$ is returned by the procedure directly as a function of $q$, in the following we will denote $D(\alpha(q))$ as $D(q)$.
The width of the MFS is the extent of the $D(\cdot)$ support, which characterizes the degree of multifractality of a series.
Clearly, all these quantities are not independent with each other and thus, in order to reduce redundancies, we only considered the subset consisting of $\alpha(q)$ and $D(q)$ in the embedding discussed later in Sec. \ref{sec:embedding}.
In fact, as said before, the MFS, $D(\cdot)$, encodes all information regarding the multifractality of the time series.
Notice that all the networks are described simultaneously by three time series, corresponding to the three observables VD, VCL, and VCC, while the probe time series are expressed by the same realization of the process for all the three observables.

\begin{figure}[ht!]
\centering

\subfigure[Time series of VD.]{
\includegraphics[viewport=0 0 347 248,scale=0.52,keepaspectratio=true]{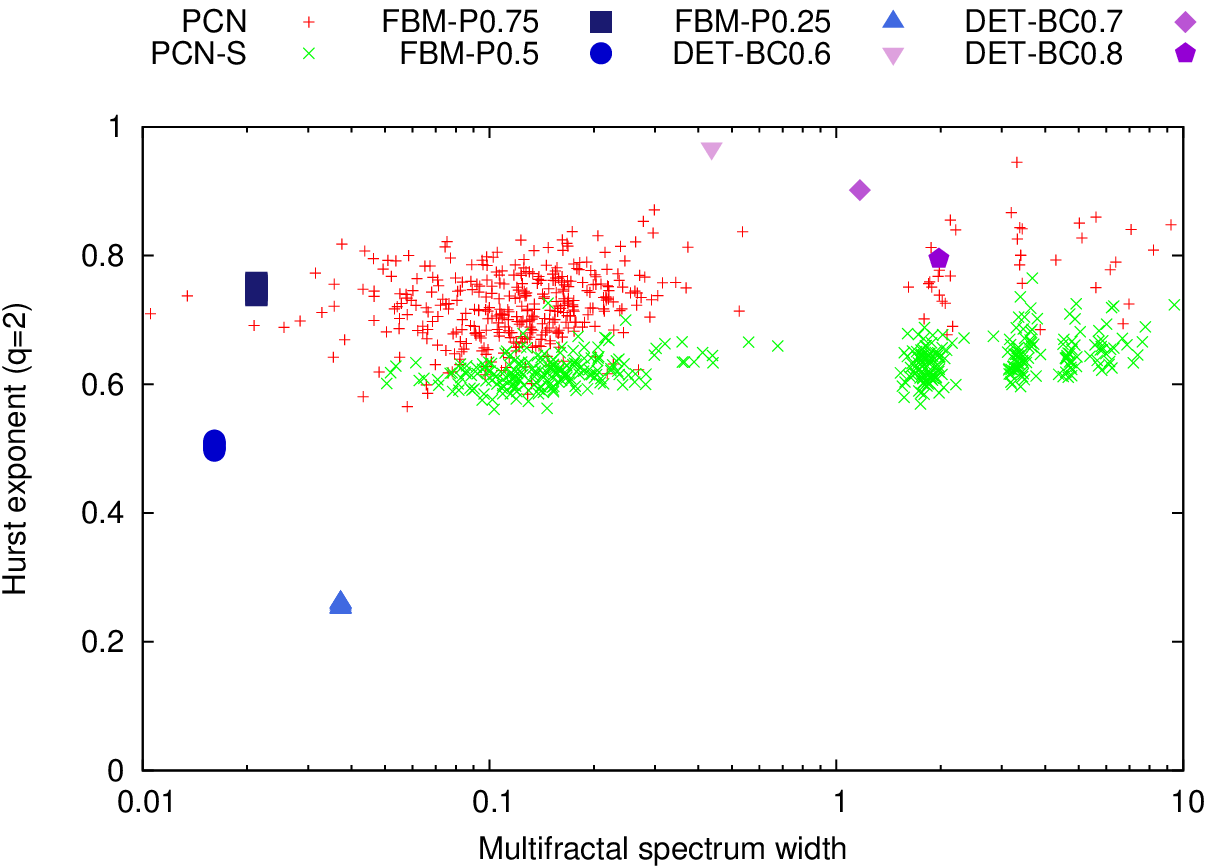}
\label{fig:spectra_deg}}
~
\subfigure[Time series of VCL.]{
\includegraphics[viewport=0 0 347 248,scale=0.52,keepaspectratio=true]{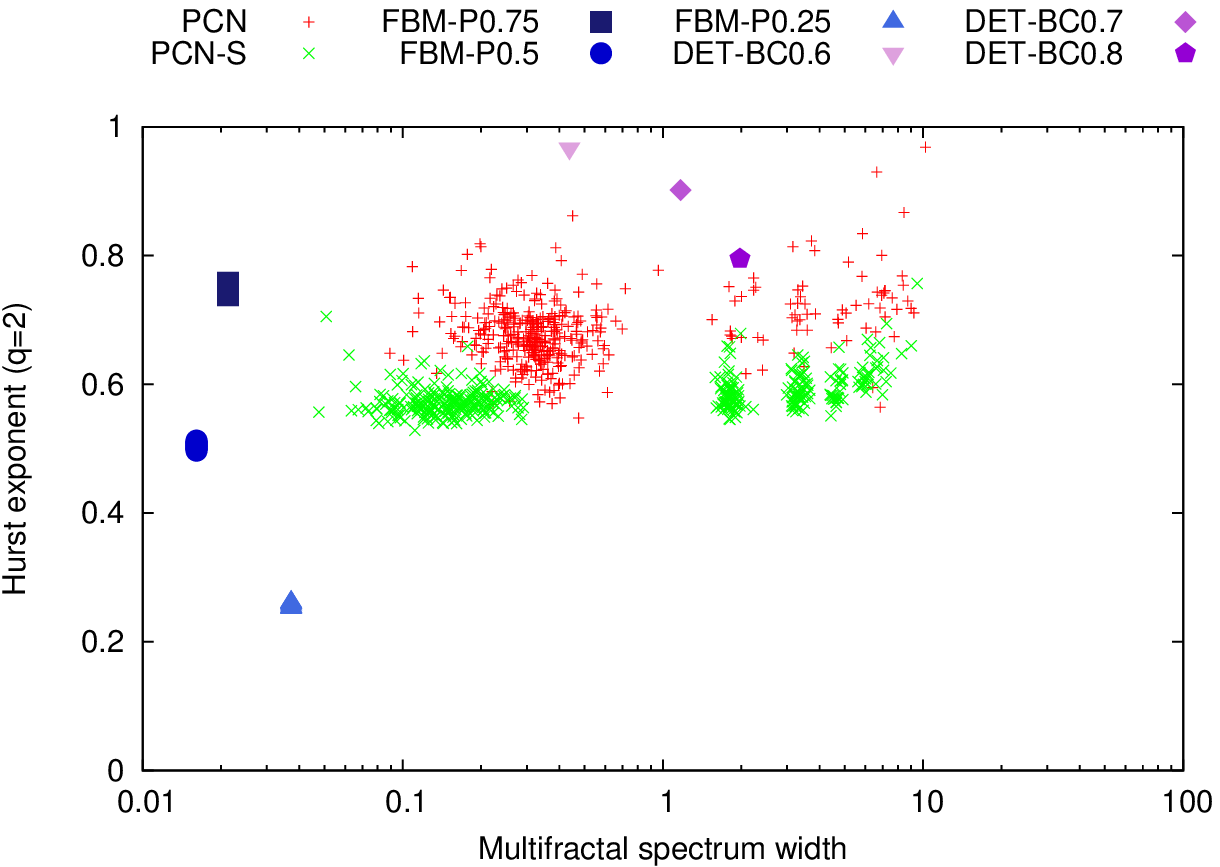}
\label{fig:spectra_cl}}

\subfigure[Time series of VCC.]{
\includegraphics[viewport=0 0 347 248,scale=0.52,keepaspectratio=true]{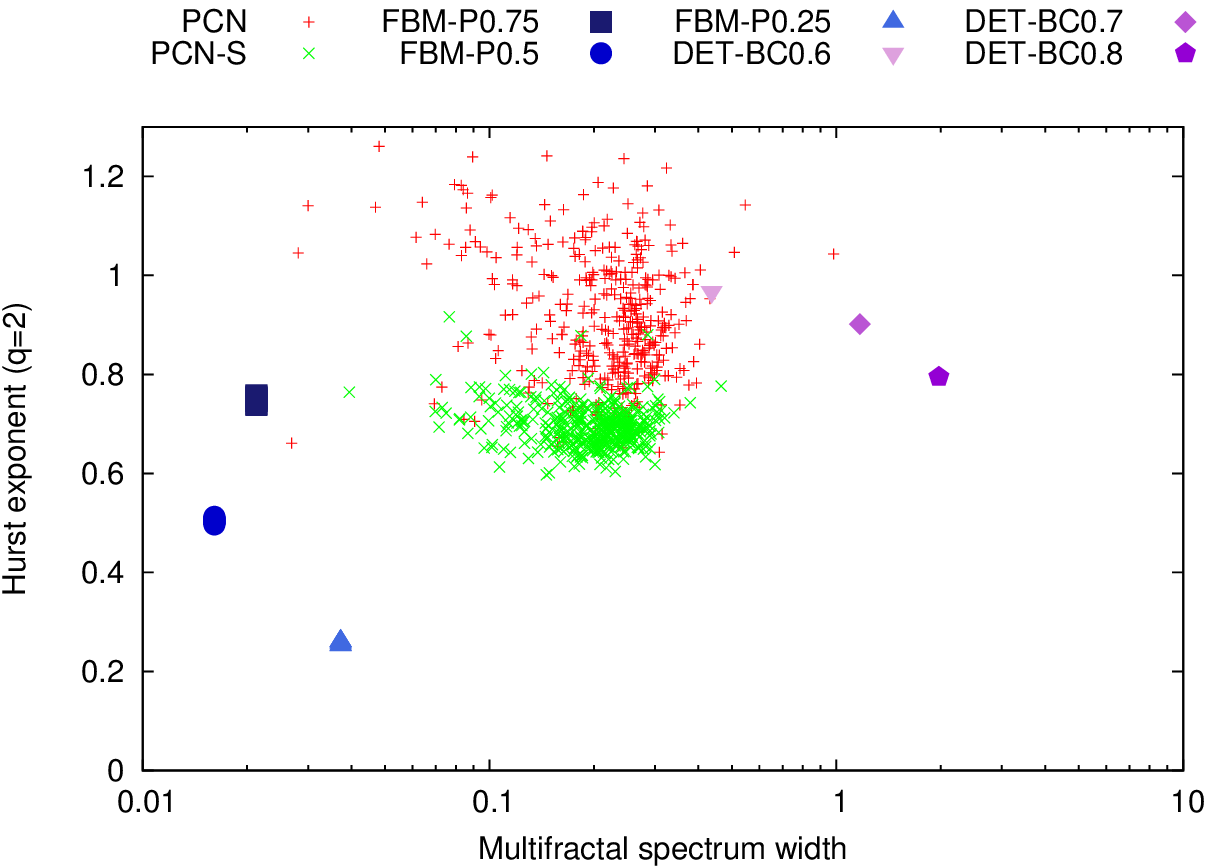}
\label{fig:spectra_cc}}

\caption{Hurst exponent vs MFS width of PCN and PCN-S. Both PCN and PCN-S show characteristics in-between mono and multi- fractal signals. Notice that the plot scale is log-lin for sake of clarity.}
\label{fig:mfs_width}
\end{figure}

To gain a first insight on the multifractality of the considered time series, it is useful to relate this property to the persistence levels calculated in Sec. \ref{sec:persistence}.
In particular, we perform this analysis for the PCN and the PCN-S since they exhibit the highest values of $H$; we consider here also the six probes, i.e., the three FBMs and three deterministic BC.
Fig.~\ref{fig:spectra_deg}, \ref{fig:spectra_cl}, and \ref{fig:spectra_cc} show the plots of $H$ versus the width of the MFS, respectively for the observables VD, VCL, and VCC.

By comparing in Fig. \ref{fig:mfs_width} the relative distances between the PCN points and the probes, it is possible to observe that most PCN exhibit MFS widths that could be considered in-between those of mono and multi-fractal signals.
Some proteins also show extremely wide MFS, while keeping the Hurst coefficient unaltered. Interestingly, PCN-S, while being less persistent, have a similar distribution of MFS widths. As observed for the persistence analysis, the VD and VCL observables behave very similarly also in terms of multifractality, while the VCC data points are more clustered and present slightly narrower spectra.
Once again, this can be attributed to the substantial difference between the types of observables.
Indeed, VD and VCL are short/medium range observables, so they can be influenced by the vertex position within the network at many distance scales. Instead, the VCC is mainly influenced by large scales (being a global topological descriptor), hence explaining why it shows less multifractal behaviour.

The variety of MFS herein observed justifies the experiments performed in the next section, which are focused on the analysis of the MFS projected in a suitable PCA space.

\subsection{Embedding of the multifractal spectra}
\label{sec:embedding}

As mentioned above, the MFS elaborated from the time series constitutes the principal hallmark of all multifractal features.
However, as first observed in Sec. \ref{sec:multifractal}, the spectra widths vary significantly even between members of the same class. Hence, there is no element that can be accounted for a meaningful representative of the whole class of proteins.
For this reason, we embed all considered MFS coefficients, i.e., $D(q)$ and $\alpha(q)$, in a suitable low-dimensional vector space derived by means of a PCA.
With the embedding into a PCA space, we are enabled to study the ensemble properties of each class without focusing on single elements alone, hence gaining an insight on the features that mostly characterize the particular typology of networks.
In such embedding, each time series is initially represented by a vector $v \in \mathbb{R}^n$. Here, $n = 300$ is the total number of coefficients retrieved by the MFDFA that we consider, which is composed by 50 values of $D(q)$ plus 50 values of $\alpha(q)$, for each of the three observables.
A given network $\mathcal{G}$, associated with time series $x^\text{VD}_\mathcal{G}(t), \, x^\text{VCL}_\mathcal{G}(t)$, and $x^\text{VCC}_\mathcal{G}(t)$, is thus represented by a vector $\vec{v}_\mathcal{G} \in \mathbb{R}^{300}$ with the form
\begin{align}
\label{eq:mfdfavec}
\begin{split}
\vec{v}_\mathcal{G}=
 &  \Big[ D_\text{VD}(-5),\,\dotsc\,, D_\text{VD}(+5),\; \alpha_\text{VD}(-5), \,\dotsc\,, \alpha_\text{VD}(+5), \Big. \\ & \;\;\;\;\;  \drsh \; D_\text{VCL}(-5),\,\dotsc\,, D_\text{VCL}(+5),\; \alpha_\text{VCL}(-5), \,\dotsc\,, \alpha_\text{VCL}(+5),  \\ & \;\;\;\;\;  \drsh \;\Big. D_\text{VCC}(-5),\,\dotsc\,, D_\text{VCC}(+5),\; \alpha_\text{VCC}(-5), \,\dotsc\,, \alpha_\text{VCC}(+5)\Big]^\top, \scriptstyle
\end{split}
\end{align}
where $D_\mathcal{O}(q)$ and $\alpha_\mathcal{O}(q)$, with $\mathcal{O} \in \{ \text{VD, VCL, VCC} \}$, are respectively the dimension coefficient and the singularity coefficient associated to the time series $x^\mathcal{O}_\mathcal{G}(t)$ as a function of the order parameter $q$.
We stress that in our analysis $q$ assumes 51 equally-spaced values between -5 and 5, with a step size of 0.2; however, we do not consider the $q=0$ case since it yields trivial values for the MFS.

On the other hand, the probe time series (FBM and BC) are not derived from a network. To be consistent with the aforementioned vector representation, their MFDFA coefficients are simply replicated 3 times, giving a vector of the form:
\begin{align}
\label{eq:mfdfa_probe}
\begin{split}
\vec{v}_\text{probe}=
 & \Big[ D(-5),\,\dotsc\,, D(+5),\; \alpha(-5), \,\dotsc\,, \alpha(+5), \Big. \\ & \;\;\;\;\;  \drsh \; D(-5),\,\dotsc\,, D(+5),\; \alpha(-5), \,\dotsc\,, \alpha(+5),  \\ & \;\;\;\;\;  \drsh \;\Big. D(-5),\,\dotsc\,, D(+5),\; \alpha(-5), \,\dotsc\,, \alpha(+5)\Big]^\top. \scriptstyle
\end{split}	
\end{align}

The 300-dimensional vector space described above is obviously unmanageable from the point of view of interpretation and, of course, visualization. For this reason, we perform a PCA to obtain a more synthetic description of the data.
The PCA does not allow only to reduce the dimensionality of the data, but it also allows to give a reasonable and more direct interpretation of the new reference framework, i.e., the PCs. This is the main reason why we opted for PCA instead of a more sophisticated, non-linear, dimensionality reduction technique. Notice also that the process has been operated on the standardized data (z-scores), which corresponds to the correlation-based PCA, instead of the covariance-based version. 

\begin{figure}[ht!]
\centering

\subfigure[PC1--PC2]{
\includegraphics[viewport=0 0 347 246,scale=0.60,keepaspectratio=true]{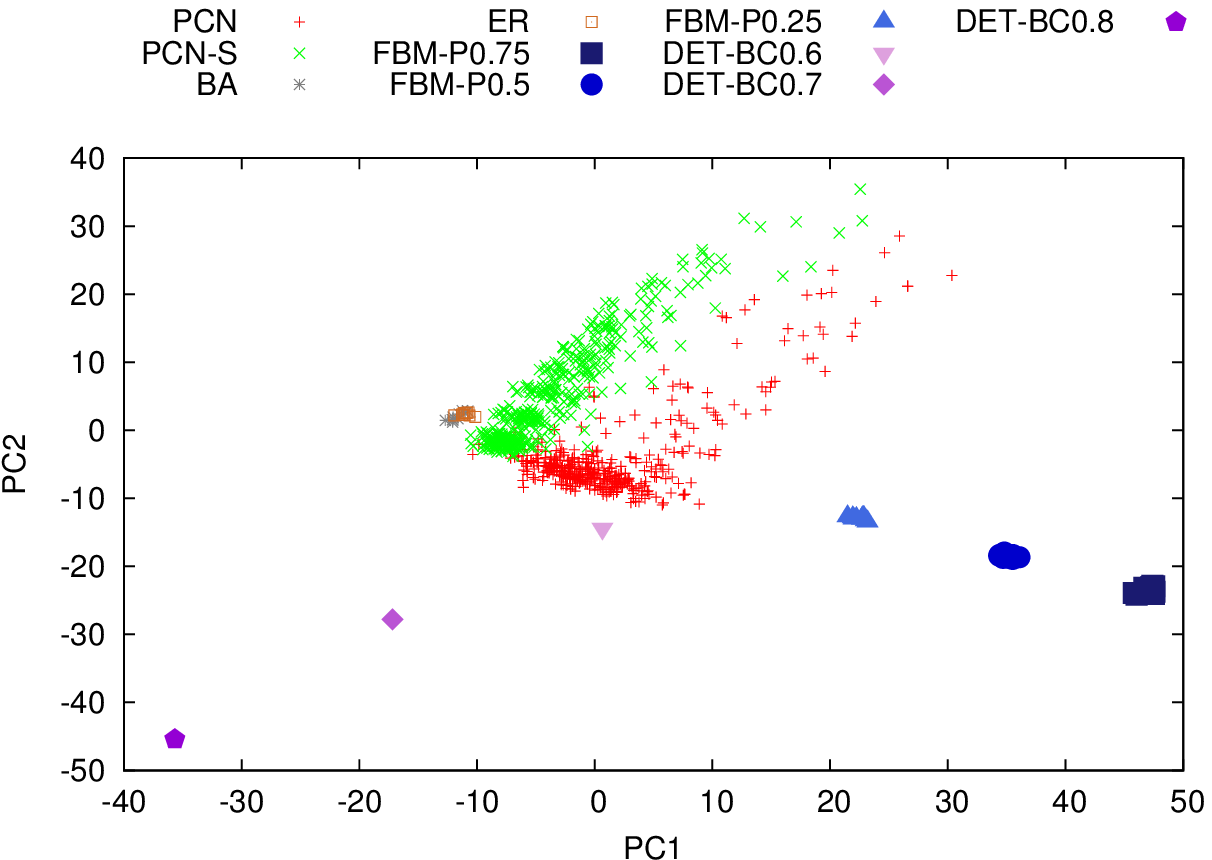}
\label{fig:mfdfa_pc1-2}}
~
\subfigure[PC3--PC4]{
\includegraphics[viewport=0 0 347 246,scale=0.60,keepaspectratio=true]{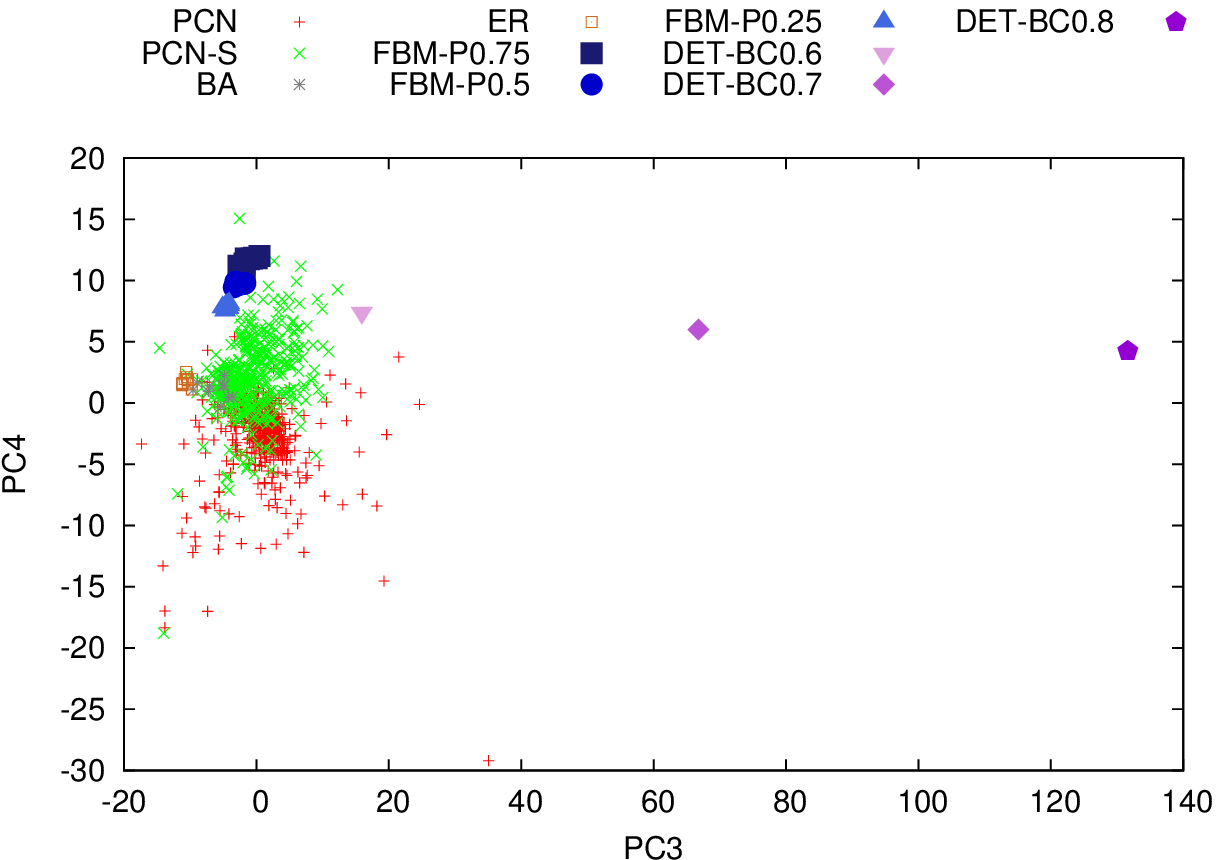}
\label{fig:mfdfa_pc3-4}}

\caption{PCA of the MFS extracted from all considered time series.}
\label{fig:PCA_MFDFA}
\end{figure}

As shown in Tab.~\ref{tbl:explvariance}, the first four PCs explain more than the 83\% of the entire variance.
For this reason, we will move our analysis to the considerably simpler four-dimensional space spanned by the first four PCs, which are shown in the plots of Fig.~\ref{fig:PCA_MFDFA}; we consider the two-dimensional subspaces derived by PC1--PC2 (\ref{fig:mfdfa_pc1-2}) and P3--PC4 (\ref{fig:mfdfa_pc3-4}), respectively.
\begin{table}
\caption{Explained variance of the first five PCs.}
\label{tbl:explvariance}
\begin{center}
\begin{tabular}{|c|c|c|c|c|c|}
\hline
& \textbf{PC1} & \textbf{PC2} & \textbf{PC3} & \textbf{PC4} & \textbf{PC5} \\ 
\hline
\textbf{Explained Variance(\%)} & 31.12 & 29.46 & 16.58 & 6.71 & 4.94 \\ 
\hline
\end{tabular}
\end{center}
\end{table}

\begin{figure}[ht!]
\centering

\subfigure[Loadings of PC1 and PC2]{
\includegraphics[viewport=0 0 350 246,scale=0.60,keepaspectratio=true]{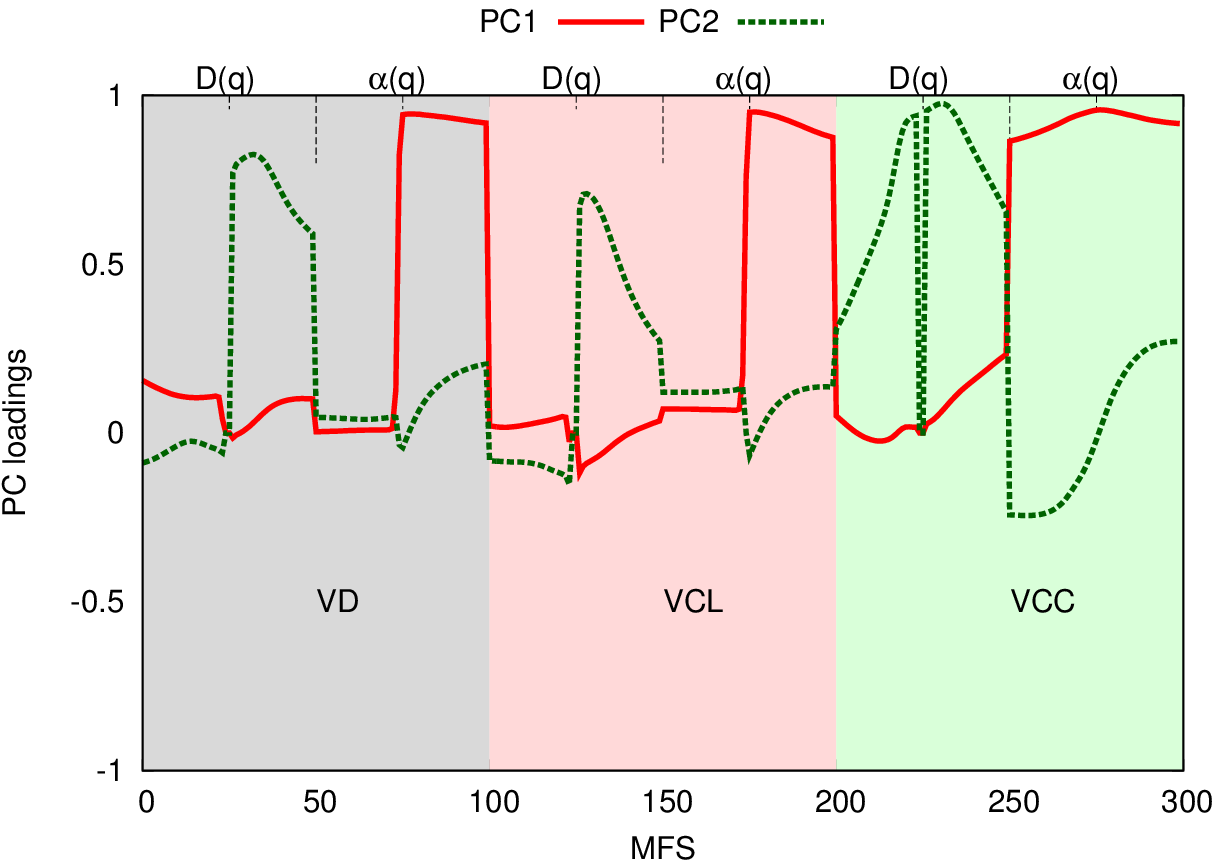}
\label{fig:fact1-2}}
\subfigure[Loadings of PC3 and PC4]{
\includegraphics[viewport=0 0 350 246,scale=0.60,keepaspectratio=true]{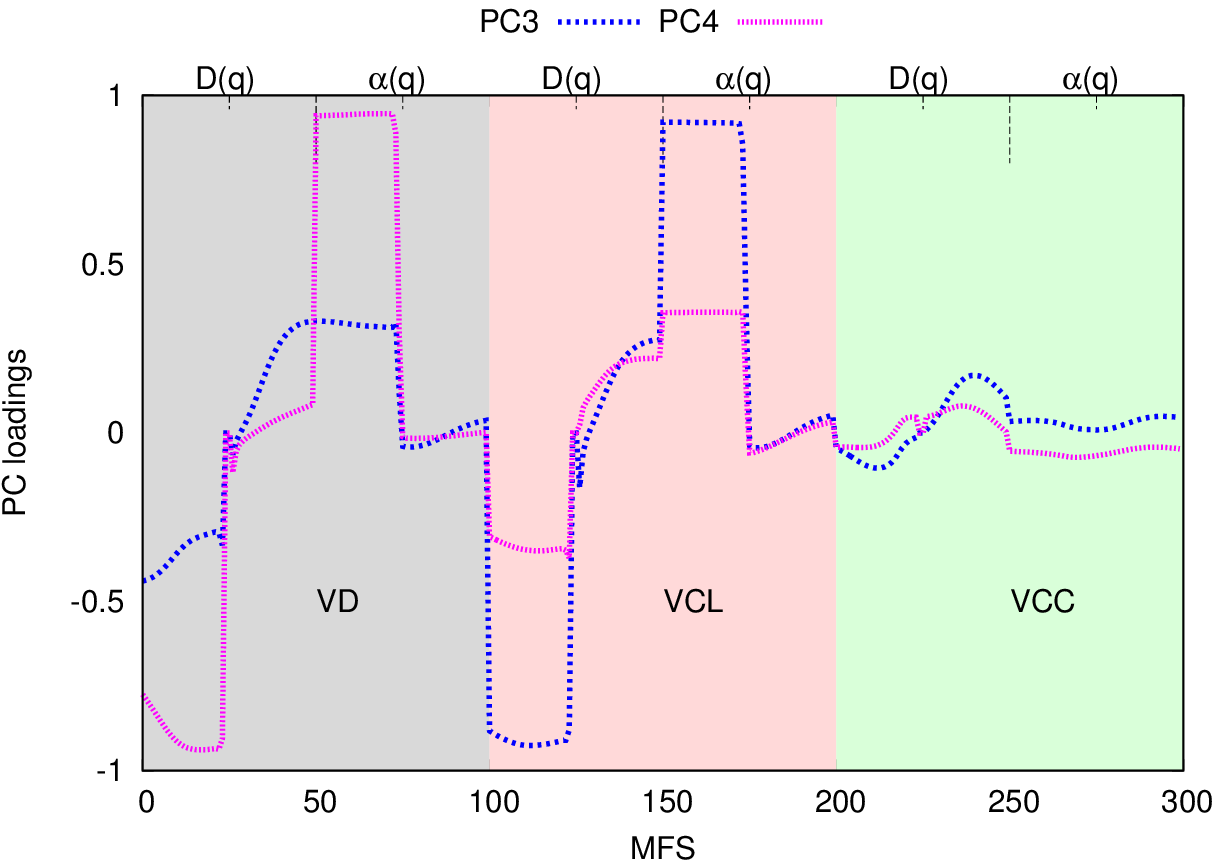}
\label{fig:fact3-4}}
\caption{PC loadings. The three observables forming the overall MFS are differentiated by using diverse shaded colors.}
\label{fig:factors}
\end{figure}

To understand the meaning of the PCs just retrieved, we analyze their loadings. In Fig.~\ref{fig:factors} are shown the correlations of each original variable with the first four PCs, where the original variables are ordered as described in Eqs. \ref{eq:mfdfavec} and \ref{eq:mfdfa_probe}.
As it is possible to note in Tab.~\ref{tbl:explvariance}, the first two principal components, PC1 and PC2, are nearly equivalent in terms of explained variance and they are correlated, respectively, to the singularity exponent $\alpha(q)$ and spectrum $D(q)$.
In particular, they are both strictly related to the large fluctuations (positive $q$ orders) of the observables VD and VCL, and to almost all fluctuation orders of VCC -- see Fig. \ref{fig:fact1-2}.
Once again, there is a clear separation between the characteristics of short and medium range observables, VD and VCL, and the long-range observable VCC. In fact, the discriminating power of VD and VCL is limited to the structure of their larger fluctuations, which is due to their local nature (we stress that local here refers to the neighborhood extent of the corresponding vertex).
Arguably, the related small fluctuations behave just as a ``background noise'', providing little information on the relevant global properties of the networks.
On the other hand, the organization of large fluctuations of VD and VCL in a RW indicates the occurrence and distribution of significant events, i.e., those related to the global topology of the network, like for example jumps between modules or areas with different local topology, hub encounters, etc.
By following this interpretation, large fluctuations provide information that appears to play an important role in the discrimination of the network's class. The VCC, instead, is fundamentally different. In this case, as mentioned at the end of Sec. \ref{sec:multifractal}, the observable is much more sensitive since it is affected by the network topology at the largest distance scales. Hence, its variations are globally discriminating at all fluctuation orders.

The loadings of the third and fourth PC shown in Fig. \ref{fig:fact3-4} are easier to interpret. In fact, the first thing that is worth noting is the complete absence of influence of the VCC observable -- since it is almost completely loaded in the first two PCs. This first observation reconfirms that all fluctuation orders of VCC provide important information in terms of variance.
Interestingly, PC3 and PC4 are suitably allocated on the VCL and VD observables. At odds with what we have observed in Fig. \ref{fig:fact1-2}, PC3 and PC4 are characteristic of the small fluctuations only (of both $D(q)$ and $\alpha(q)$). 
However, PC3 seems to be heavily influenced by the probe networks variance; Fig. \ref{fig:mfdfa_pc3-4} offers a visual understanding of this claim. Therefore, its contribution in the discrimination among the different network topologies is questionable, while PC4 provides a small but yet perceptible contribution in terms of variance.

\section{Conclusions}
\label{sec:conclusions}

In this paper, we have exploited the possibility to characterize protein contact networks by means of a multifractal analysis of suitable time series. Such time series have been generated by performing stationary, unbiased, random walks on the graph structures, recording at each vertex three different quantities: the degree, clustering coefficient, and closeness centrality of the vertex.
Those three observables capture, respectively, short, medium, and long range peculiarities of the considered networks.
Our analysis of the considered protein contact networks was compared with several probe data. Notably, we used a receipt to generate synthetic polymers designed to mimic random coiled cords, two well-known classes of random networks, and two models of time series embodying the archetypical monofractal and multifractal signals.

The herein presented study provided a number of results. First, persistence analysis of the time series showed that proteins, regardless of the considered vertex observable, generate strongly persistent signals.
When considering the degree as the observable, this can be translated into assortativity of the degree distribution. This first result is confirmed by the recent literature, although, to our knowledge, we are the first to asses such a property by means of time series analysis. We also pointed out that this result is in contrast with the recent hypothesis requiring disassortativity in fractal networks \cite{song2006origins,gallos2007review,zhang2007self}.
We also found that the assortativity of other observables can be linked to the assortativity of the vertex degree, since the degree basically controls the behaviour of the RW and thus influences to some extent all other measurements.
Then we moved to a first analysis inspecting the multifractal footprint proper of the considered time series.
Results showed that time series associated to protein contact networks -- again regardless of the observable -- should be considered as signals in-between the typical mono and multi- fractal behavior.
We further elaborated over those results by performing the interpretation of the entire multifractal spectrum via the embedding into a suitable vector space. Such a vector space has been derived by first associating each time series to a high-dimensional vector containing suitable samplings of the domain and codomain of the multifractal spectrum derived by the multifractal dentrended fluctuation analysis. Successively, we performed a principal component analysis, resulting in a four-dimensional vector space explaining large part of the original data variance.
The principal component analysis allowed us to perform a detailed interpretation regarding the importance of different fluctuation orders by analysing their loadings on the principal components.
Results showed that large (in magnitude) fluctuations of all observables are more important in terms of discrimination (variance) of the considered networks/time series. Along with these, small fluctuations of the closeness centrality observable were also recognized to be discriminating, fact that has been attributed to their long-range (global) nature. Small fluctuations of the degree and clustering coefficient, instead, are less informative, since they are more easily associated with background noise.

We conclude by arguing that the herein presented study for analyzing complex networks could be used also in different settings. Indeed, the techniques employed in this work never assume the knowledge of the global topology of the graph. In particular, this study might be of interest when the topology of the network under analysis is not directly observable, but can be gradually ``explored'' with suitable time-dependent measurements of the vertices.
Moreover, the comparison of the herein considered proteins with the corresponding synthetic versions highlighted important differences, which in turn strengthen the need to develop a more suitable generative model for protein contact networks. Such a generative model, if able to reproduce the co-existence of fast-lane communication by specific long-range contacts with the strong modularity of protein contact networks, could be of utmost importance in a number of technological applications.

\bibliographystyle{abbrvnat}
\bibliography{Bibliography.bib}
\end{document}